\documentclass[slac_one]{revtex4}
\usepackage{graphicx}
\usepackage{fancyhdr}
\begin{document}

\title{{\small{Hadron Collider Physics Symposium (HCP2008),
Galena, Illinois, USA}}\\ 
\vspace{12pt}
Search for $CP$ Violation in $B^0_s\rightarrow J/\psi \phi$ at CDF} 

%

\author{Chunlei Liu}
\affiliation{University of Pittsburgh, Pittsburgh, USA}

\begin{abstract}
The CKM mechanism is well established as the dominant mechanism for $CP$ violation, which was first discovered in 
the neutral kaons  in 1964~\cite{ref:cpkaon}. 
To search for new sources of $CP$ violation, one can exploit a handful of systems in which the standard model makes
a precise prediction of $CP$ violation. In the $B^0_s\rightarrow J/\psi \phi$ system, $CP$ violation in the interference
of mixing and decay is precisely predicted in the standard model, the prediction being very close to zero. The CDF experiment
reconstructs about 2000 signal events in 1.35~fb$^{-1}$ of luminosity. We obtain a confidence region in the space of the
parameters $\beta_s$, the $CP$ phase, and $\Delta\Gamma$, the width difference. This result is 1.5~$\sigma$ from the standard
model prediction.
\end{abstract}

\maketitle

\thispagestyle{fancy}


\section{INTRODUCTION}
\begin{figure*}[htb]
\centering
\makebox{
\includegraphics[width=60mm]{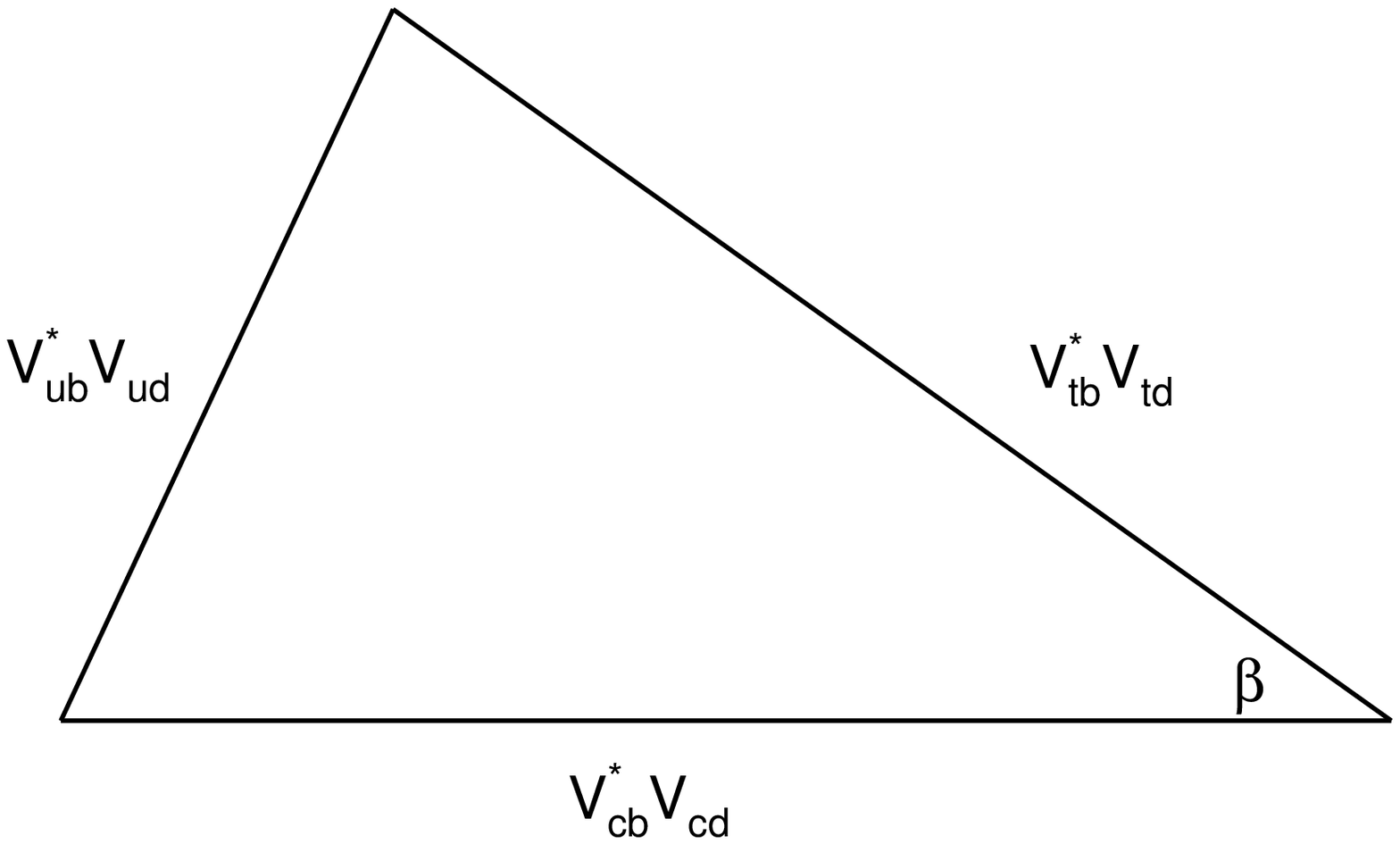}
\includegraphics[width=60mm]{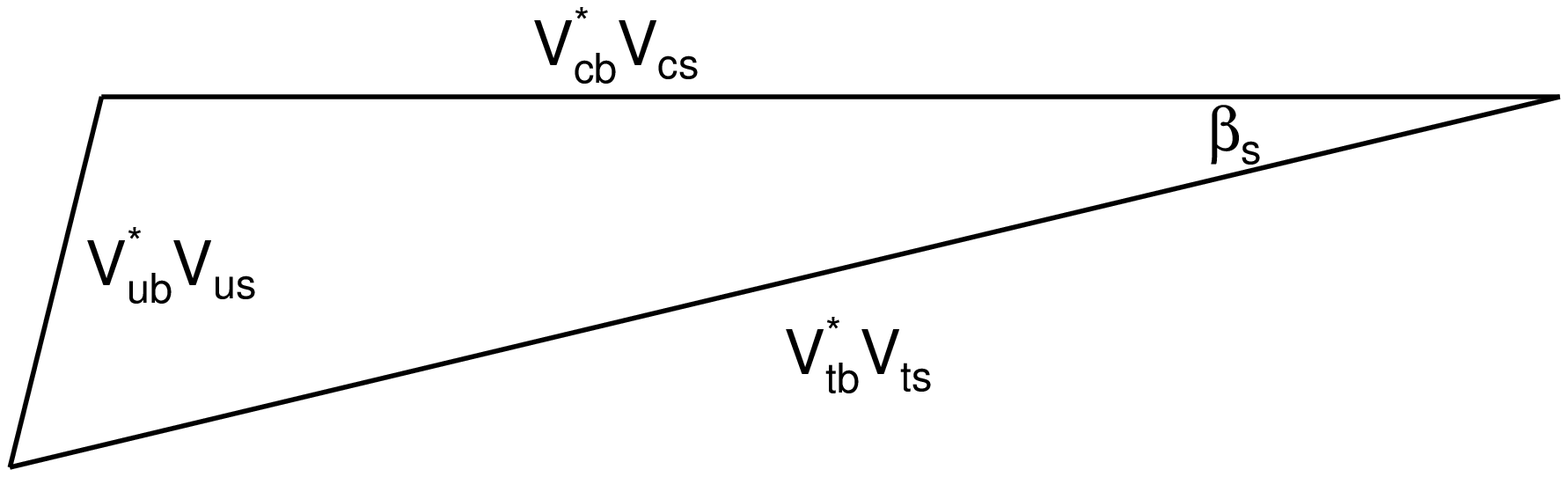}}
\caption{Unitarity triangles from CKM matrix.} 
\label{fig:triangle}
\end{figure*}

$CP$ violation in the standard model is associated with the CKM matrix~\cite{ref:CKM1,ref:CKM2}, which arises from the 
charged $W$ transition. Three generations of quarks lead to a $3\times3$ unitary matrix $V_{\rm CKM}$ 
with four independent parameters: three mixing angles and one imaginary phase, and  
the phase is the source of $CP$ violation in the standard model.
The Wolfenstein parametrization of the CKM matrix is useful~\cite{ref:Wolf} 
\begin{eqnarray}
V_{\rm CKM}  =  \left( \begin{array}{lcr}
  V_{ud}   & V_{us}   & V_{ub}   \\
  V_{cd}   & V_{cs}   & V_{cb}   \\
  V_{td}   & V_{ts}   & V_{tb}
 \end{array} \right)  = 
 \left( \begin{array}{ccc}
  1 - \frac{1}{2}\lambda^2- \frac{1}{8}\lambda^4   & \lambda        & A \lambda^3(\rho - i \eta)   \\
  -\lambda+\frac{1}{2}A^2\lambda^5-A^2\lambda^5(\rho+i\eta)    & 1-\frac{1}{2}\lambda^2-\frac{1}{8}\lambda^4(1+4A^2)  & A\lambda^2   \\
  A\lambda^3(1 - \bar{\rho} - i\bar{\eta})  & -A \lambda^2+A\lambda^4(\frac{1}{2}-\rho-i\eta)   & 1-\frac{1}{2}A^2\lambda^4
 \end{array} \right)
\end{eqnarray}
The unitarity property of the CKM matrix gives six unitarity triangles. For example, if we apply 
unitarity to the first and third columns, we get the following equation
\begin{equation}
V^*_{ub}V_{ud}+ V^*_{cb}V_{cd}+V^*_{tb}V_{td}=0
\end{equation}
which can be represented as a triangle in the complex plane as shown in 
Fig.~\ref{fig:triangle} (left). The CKM mechanism predicts a sizable angle 
$\beta\equiv {\rm arg} \left(-\frac{V_{cd}V^*_{cb}}{V_{td}V^*_{tb}}\right),$ 
since all the three sides are at the same order of $\lambda^2$. This angle can be cleanly measured in 
the $B^0\rightarrow J/\psi K^0_s$ decay through the time dependent $CP$ asymmetry. The large 
angle is an indication of large $CP$ violation in the $B^0$ systems, which was verified in 2001~\cite{ref:b0cp1,ref:b0cp2}.
If unitarity is applied to the second and third
columns, another equation is obtained
 \begin{equation}
V^*_{ub}V_{us}+ V^*_{cb}V_{cs}+V^*_{tb}V_{ts}=0
\end{equation}
which also corresponds to a triangle in the complex plane as shown in Fig.~\ref{fig:triangle} (right). 
However, two sides of this triangle are at order of $\lambda^2$, and the third side is only at order of $\lambda^4$. 
This leads to a very small angle ($\sim \lambda^2$) defined as
$\beta_s\equiv {\rm arg} \left(-\frac{V_{ts}V^*_{tb}}{V_{cs}V^*_{cb}}\right).$ 
This angle can be cleanly measured in the decay $B^0_s \rightarrow J/\psi \phi$.
The measurement of the $CP$ violation phase $\beta_s$
is thus very interesting and any significant non-zero result could be an indication of new physics beyond the standard model.
Eigenstates of the full Hamiltonian $H$, the mass eigenstates $B^L_s$ and $B^H_s$~\cite{ref:mixing} are  $CP$ eigenstates if 
$[H,CP]=0$. 

In these proceedings, we will present the first measurement of $CP$ violation in $B^0_s\rightarrow J/\psi \phi$
decays with flavor identification. The events are collected at CDF experiment located at Fermilab. 
The details about the experiment can be found in Ref.~\cite{ref:CDFd}. 

\section{EVENT RECONSTRUCTION}

\begin{figure*}[htb]
\centering
\makebox{
\includegraphics[width=57mm]{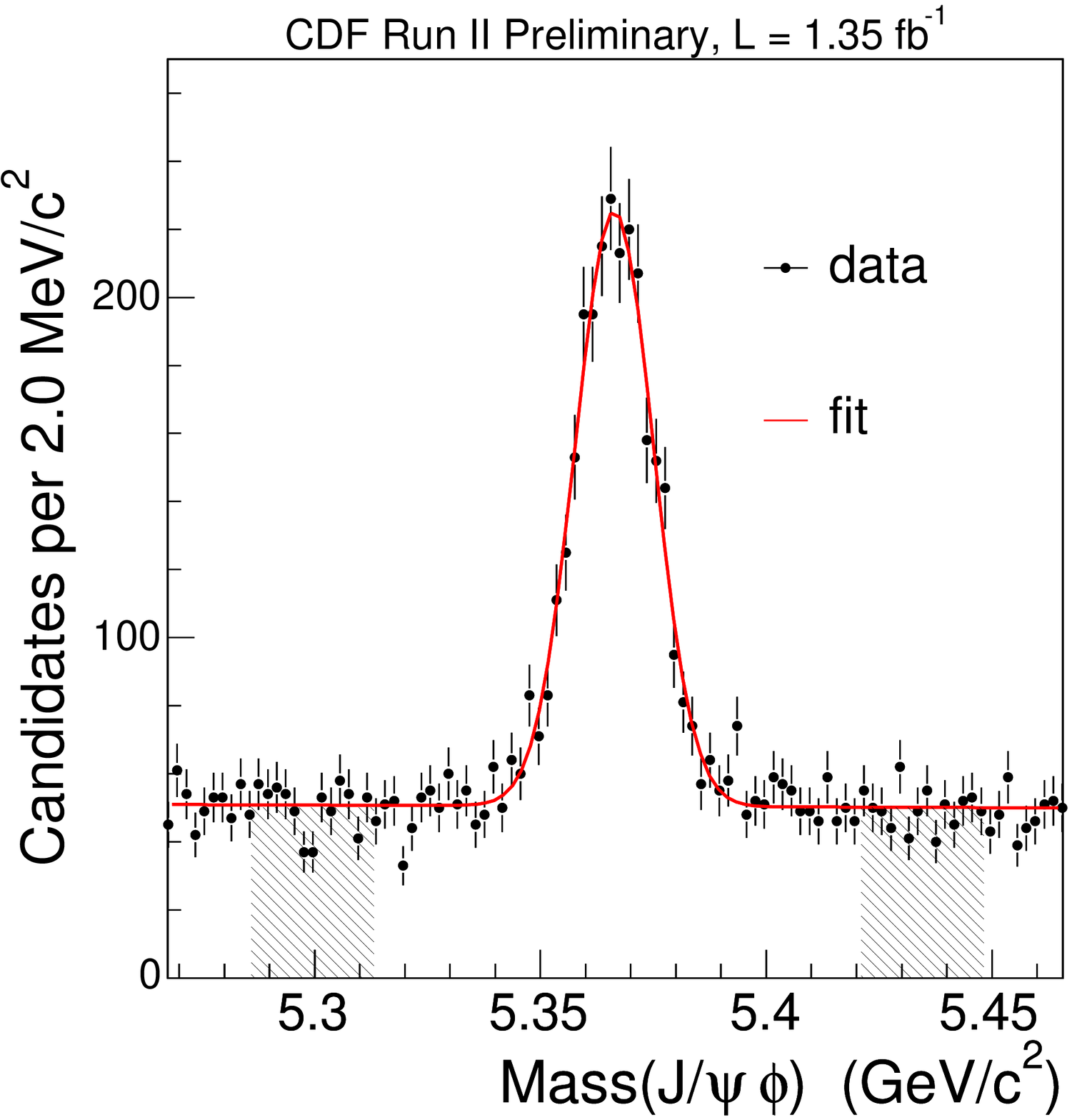}
\includegraphics[width=64mm]{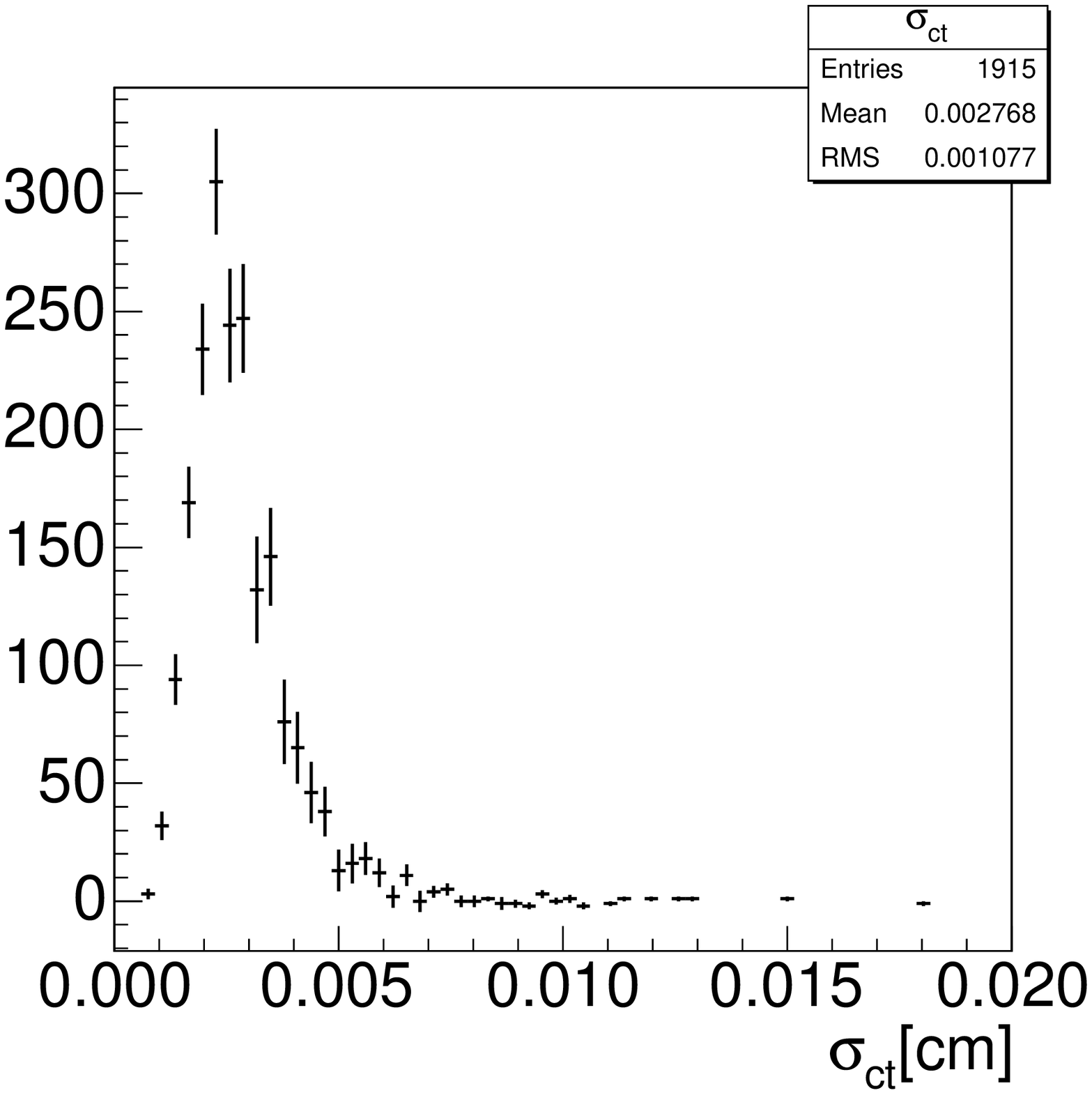}}
\caption{(left) $B^0_s$ invariant mass distribution, (right) sideband subtracted signal proper decay length uncertainty distribution.} 
\label{fig:mass}
\end{figure*}

The data are collected with a dedicated di-muon trigger at CDF, which preferentially chooses $J/\psi$ mesons
that decay to two muons. A $J/\psi$ candidate is reconstructed from
two muon tracks of opposite charges and the reconstructed mass is within a 80~MeV wide mass window about the PDG value.
A $\phi$ candidate is obtained from two oppositely charged non-muon tracks with reconstructed mass within a 12~MeV wide mass
window about the PDG value. A $B^0_s$ vertex is formed from all the daughter tracks. After some loose pre-selection 
(transverse momentum cut), the data are selected by an artificial neural network (NN).
The network needs to be trained before being applied to the data. To separate signal and combinatorial background events, 
both signal and background samples are provided to the neural network for the training. The signal sample is obtained from 
Monte Carlo simulation, while the background sample is 
obtained from the $B^0_s$ mass sideband region. The variables used for training  mainly include: vertex fit probability
at each vertex, transverse momentum of $B^0_s$ and daughter tracks, and particle identification information for kaon candidates.
The neural network assigns a numerical output value for each event in the data sample, and a NN output cut is 
chosen to optimize the figure of merit: $S/\sqrt{S+B}$, where $S$ and $B$ are number of signal and background events
in the defined signal mass region. With 1.35~fb$^{-1}$ of luminosity, we get about 2000 signal events. The invariant
$B^0_s$ mass distribution of events after NN selection is shown Fig.~\ref{fig:mass} (left).

\section{EXPERIMENTAL STRATEGIES}

To observe any time dependent $CP$ asymmetry, we need to measure the decay rates of both  $B^0_s$ and $\bar{B}^0_s$.
This requires proper decay time measurement with excellent resolution in order to resolve the fast oscillation of the differential
rates, and the measurement of final state decay angles to separate different $CP$ eigenstates. 
In addition, we need to develop algorithms to identify the flavor of the $B^0_s$ meson at the production time.   

\subsection{Proper Decay Time and Uncertainty}

The distance $L$ between the primary vertex ($B^0_s$ produced) and the secondary vertex ($B^0_s$ decays) is the 
decay length measured in the lab frame. The primary vertex is reconstructed on an event by event basis with 
average uncertainty around $30~\mu m$. To the get decay time in the $B^0_s$ rest frame, the projected decay length  
on the $x-y$ plane (perpendicular to the beamline) is used. The proper decay length is
\begin{equation}
ct= \frac{L_{xy}M}{p_T}
\end{equation}  
where $c$ is the light speed, $M$ is $B^0_s$ PDG mass and $p_T$ is the reconstructed $B^0_s$ transverse momentum.
With large drift chamber and silicon vertex tracking system close to the beamline, the $ct$ resolution 
is excellent at CDF ($\sim 20~\mu m$), which can be seen  in Fig.~\ref{fig:mass} (right).

\subsection{$CP$ Eigenstates Separation}

\begin{figure*}[htb]
\centering
\includegraphics[width=80mm]{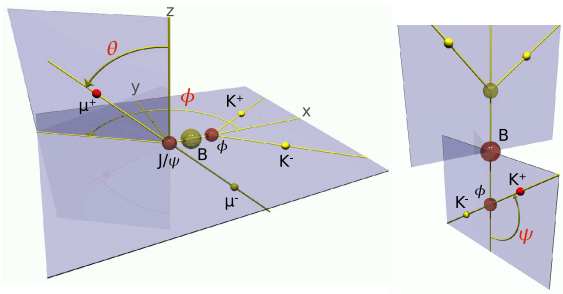}
\caption{(left) The $J/\psi$ rest frame, where direction of the $\phi$ defines $x$ axis and the plane of the $K^+K^-$ system defines $y$
axis with $p_y(K^+)>0$. (right) The $\phi$ rest frame, where $\psi$ is the angle between $K^+$ and the negative of the direction of 
$J/\psi$ in that rest frame.} 
\label{fig:angle}
\end{figure*}

Since $B^0_s$ is a pseudo-scalar meson, $J/\psi$ and $\phi$ are both vector mesons, the final states 
form a admixture of $CP$ eigenstates, where $S$ and $D$ waves are $CP$ even, while $P$ wave
is $CP$ odd. To separate the two $CP$ eigenstates, we analyze the angular distribution in the transversity
basis. 
In this basis, the final state consists of three orthogonal polarization states, such that the two final state
vector mesons are either longitudinally polarized, transversely polarized and perpendicular to each other, or
transversely polarized and parallel to each other. Transitions from the initial state to these polarization 
states are described by amplitudes ($A_0$,\ $A_{\perp}$,\ $A_{\parallel}$). 
Three transversity angles defined in different rest frames as shown in Fig.~\ref{fig:angle}
are the polar and azimuthal angles of positive muon in the $J/\psi$ rest frame, 
and the helicity angle of positive kaon in the $\phi$ rest frame. 

\subsection{Flavor Identification}

$b$ and $\bar{b}$ quarks are produced together in general through QCD at Tevatron. Two types of tagging 
algorithms are used at CDF. The first algorithm tags the $b$ quark that produces the $B^0_s$ candidate
in the sample, which is called same side tagging (SST). The other algorithm, known as opposite side
tagging (OST), tags the other quark. On the near side, the $B^0_s$ is usually associated with a charged kaon due to 
the fragmentation process, so the charge of the kaon can be used to identify  the $B^0_s$ flavor. This algorithm is also 
called same side kaon tagging (SSKT). On the away side, the charge of leptons coming from semileptonic 
decay of $B$ hadrons, or the charge of the $b$ jet is correlated to the $B^0_s$ flavor. 

The flavor tagging algorithms have limited analyzing power for several reasons. On the near side, a charged kaon 
is not always available, and has a background from charged pions. On the away 
side, both the oscillation of neutral $B$ mesons and the sequential decay of $b$ quark lead to incorrect 
flavor identification. Each tagging algorithm returns two things: 1) A decision ($\xi=\pm1, 0$) that identifies
the flavor of the $B^0_s$ candidate. The fraction of the events with decision $|\xi|=1$ is called the tagging 
efficiency $\epsilon$. 2) A quality estimate of that decision, called dilution ${\cal D}$. The probability to 
obtain a correct tag is $(1+{\cal D})/2$. The OST efficiency is around 96\%, with an average dilution around 11\%, while the SSKT 
efficiency is around 51\%, with an average dilution around 27\%. The total effective tagging power is characterized by 
$\epsilon {\cal D}^2 \sim 4.8\%$.

\section{$\beta_s$ MEASUREMENT WITH FLAVORING TAGGING}
The decay of the $B^0_s$ meson depends on both time and angular distribution, and the decay probability density function for 
$B^0_s$ can be expressed as
\begin{eqnarray}
 \frac{d^4P(t,\vec{\rho})}{dtd\vec{\rho}} & \propto & |A_0|^2{\cal T_+} f_1(\vec{\rho})+ |A_{\parallel}|^2{\cal T_+} f_2(\vec{\rho}) 
      + |A_{\perp}|^2{\cal T_-} f_3(\vec{\rho}) \nonumber \\
    &   & + |A_{0}||A_{\parallel}|\cos(\phi_{\parallel}){\cal T_+}f_4(\vec{\rho})
                    +    |A_{\parallel}||A_{\perp}|{\cal U_+} f_5(\vec{\rho}) 
                    +    |A_{0}||A_{\perp}| {\cal V_+} f_6(\vec{\rho}) \label{eqn:Rcomp}
\end{eqnarray}
where $\vec{\rho}=(\cos\theta,\phi,\cos\psi)$, and functions $f_1(\vec{\rho})\ldots f_6(\vec{\rho})$ are related to angular 
distribution. The probability density function for $\bar{B}^0_s$ is obtained by substituting ${\cal U}_+\rightarrow {\cal U}_-$
and ${\cal V}_+\rightarrow {\cal V}_-$. The time dependent term ${\cal T}$ is defined as
\begin{eqnarray}
{\cal T}_{\pm}  =  e^{-\Gamma t}  \times  \left[ \cosh{\frac{\Delta \Gamma}{2}t} \mp
                       \cos{2\beta_s}\sinh{\frac{\Delta \Gamma}{2}t} \mp \eta {\sin{2\beta_s} \sin{\Delta m t}} \right] \nonumber
\end{eqnarray}
where $\eta=+1$ for $B^0_s$ and $-1$ for $\bar{B}^0_s$. Other time independent terms are
\begin{eqnarray}
{\cal U}_{\pm}  =   \pm e^{-\Gamma t} & \times & \left[ \sin(\phi_{\perp}-\phi_{\parallel})\cos(\Delta mt)  
                                        -\cos(\phi_{\perp}-\phi_{\parallel})\cos(2\beta_s)\sin(\Delta mt) \right. \nonumber \\
                                   &  \pm  & \left. \cos(\phi_{\perp}-\phi_{\parallel})\sin(2\beta_s)\sinh(\frac{\Delta\Gamma t}{2}) \right]
                                           \nonumber \\
{\cal V}_{\pm}  =   \pm e^{-\Gamma t} & \times & \left[\sin(\phi_{\perp})\cos(\Delta mt)  
                                        -\cos(\phi_{\perp})\cos(2\beta_s)\sin(\Delta mt) \right. \nonumber \\
                              &   \pm & \left. \cos(\phi_{\perp})\sin(2\beta_s)\sinh(\frac{\Delta\Gamma t}{2}) \right]  \nonumber
\end{eqnarray}
where $\phi_{\parallel}\equiv {\rm arg}(A^*_{\parallel}A_0)$, $\phi_{\perp}\equiv {\rm arg}(A^*_{\perp}A_0)$, and $\Delta m$
is the mass difference of the two $B^0_s$ mass eigenstates which will be constrained to the CDF measurement result~\cite{ref:deltam}. 
An unbinned maximum likelihood fit is performed to extract the parameters of interest: $CP$ violation phase $\beta_s$ and decay
width difference $\Delta\Gamma$ of the two $B^0_s$ mass eigenstates. The resolution effects and detector efficiencies are also incorporated
into the likelihood function~\cite{ref:betas-tagging}.

\begin{figure*}[htb]
\centering
\makebox{
\includegraphics[width=60mm]{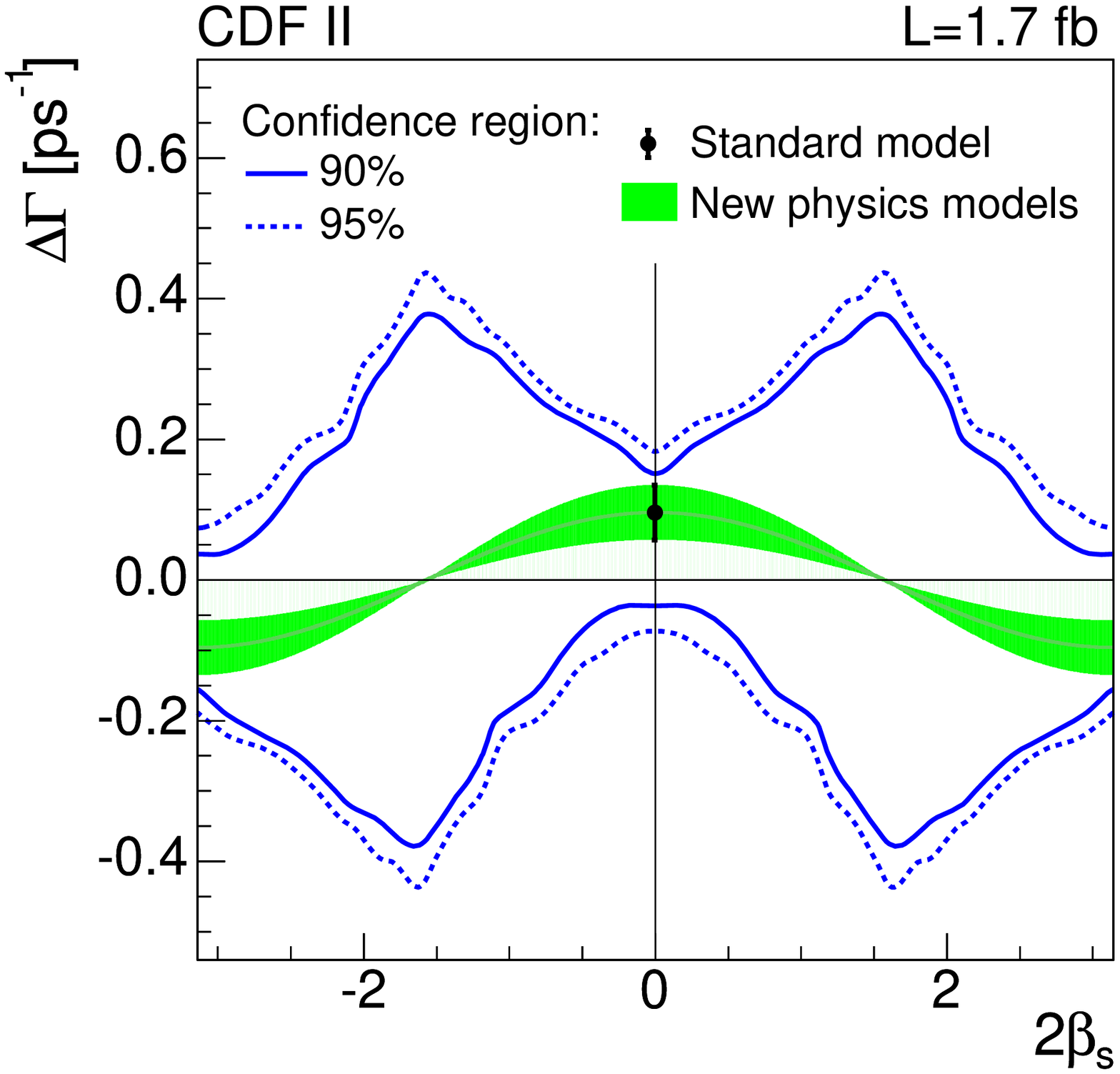}
\includegraphics[width=60mm]{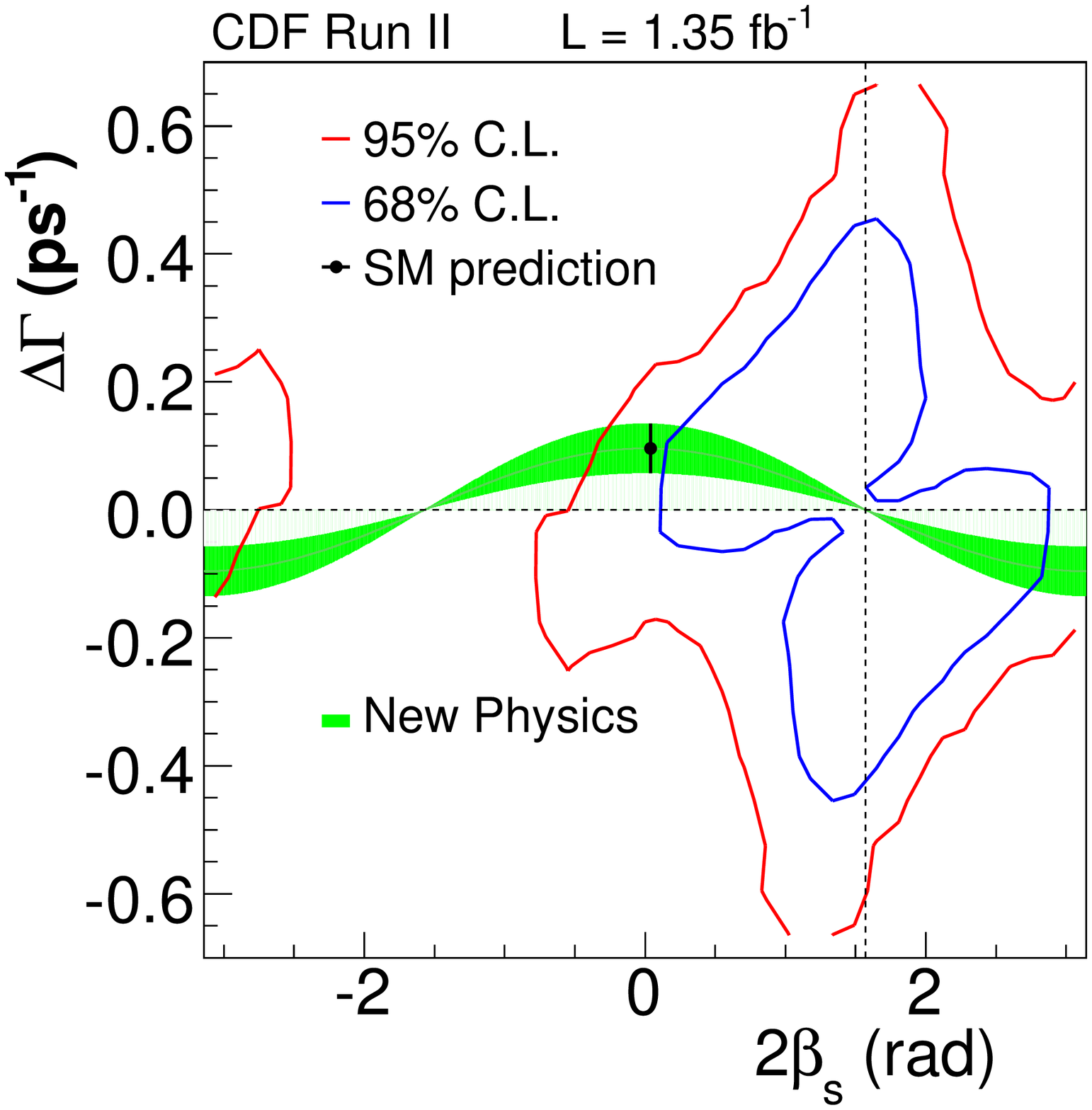}}
\caption{(left) Two dimensional confidence region of $2\beta_s$ and $\Delta\Gamma$ without flavoring tagging. 
(right) Confidence region with flavor tagging, where space of solution is reduced 50\% compared with the result 
without flavor tagging.} 
\label{fig:confidence}
\end{figure*}

Without flavor tagging, a four-fold ambiguity will arise from the likelihood function, and the 
sensitivity to $\beta_s$ is marginal. At CDF, the $\beta_s$ measurement without flavor tagging is performed with integrated 
luminosity of 1.7~fb$^{-1}$. The final result as shown in the Fig.~\ref{fig:confidence} (left) is a two dimensional confidence 
region of $2\beta_s$ and $\Delta\Gamma$. The standard model prediction of $(2\beta_s,\Delta\Gamma)\sim (0.0,0.1~{\rm ps}^{-1})$~\cite{ref:SM} 
is consistent with the data at probability of 22\% or 1.2~$\sigma$ level. The details of the measurement can be found in 
Ref.~\cite{ref:betas-untagging}.

The $\beta_s$ measurement can be improved  with flavor tagging, where one will expect better $\beta_s$ sensitivity, since we
obtain information on $B^0_s$ and $\bar{B}^0_s$ separately. However, a two-fold ambiguity still remains with the 
simultaneous transformation ($2\beta_s\rightarrow \pi-2\beta_s, \Delta\Gamma\rightarrow -\Delta\Gamma, \phi_{\parallel} 
\rightarrow 2\pi-\phi_{\parallel}, \phi_{\perp}\rightarrow \pi-\phi_{\perp}$). This symmetry, combined with limited statistics,
precludes a point estimate of the physics parameters $\beta_s$ and $\Delta\Gamma$; instead, a confidence region  
is obtained. The two dimensional confidence region of $\beta_s$ and $\Delta\Gamma$
is shown in the  Fig.~\ref{fig:confidence} (right), where the standard model prediction point $(2\beta_s,\Delta\Gamma)= 
(0.04,0.096~{\rm ps}^{-1})$ has probability 15\%, equivalent to 1.5 Gaussian standard deviation.  If $\Delta\Gamma$
is treated as a nuisance parameter, we obtain a one dimensional confidence region of $\beta_s$, where $2\beta_s \in [0.32, 2.82]$
at 68\% confidence level.

\section{CONCLUSION}

The first $CP$ violation phase $\beta_s$ measurement from flavor tagged $B^0_s\rightarrow J/\psi \phi$ decays has been presented.
The result is consistent with the standard model, but only at 15\% confidence level. Since the HCP conference, CDF has updated its
result to 2.8~fb$^{-1}$ of data which shows the result is consistent with the standard model only at 7\% confidence 
level~\cite{ref:betasupdate}.

\begin{acknowledgments}
I would like to thank the organizers of HCP 2008 conference for a enjoyable week. I also would like to thank all my colleagues at 
CDF collaborations and Fermilab staff for their hard work which makes these results possible. 
\end{acknowledgments}

\end{document}